\documentclass{iopconfser}
\usepackage{aas_macros}
\usepackage[T1]{fontenc}
\usepackage{lmodern}
\usepackage{color}
\usepackage{amsfonts,amssymb,amsmath}
\usepackage[citecolor = db]{hyperref}
\hypersetup{colorlinks=true}
\usepackage{mathtools}
\usepackage{tikz}
\usepackage{float}
\usepackage{orcidlink}
\usepackage[dvipsnames]{xcolor}
\usepackage[utf8]{inputenc}
\usepackage{cleveref}
 \crefname{section}{Section}{Sections}
 \crefname{equation}{Eq.}{Eqs.} 
 \crefname{figure}{Fig.}{Figs.}
\definecolor{db}{rgb}{0.0, 0.0, 0.62}
\definecolor{dm}{rgb}{0.7, 0.01, 0.7}
\definecolor{dr}{rgb}{0.55, 0.0, 0.0}

\begin{document}

\title{Dark matter halos and transonic accretion flow}

\author{Avijit Chowdhury\,\orcidlink{0000-0002-7235-5076}$^{1}$, Gargi Sen\,\orcidlink{0000-0002-7929-1634}$^{1}$, Sayan Chakrabarti\,\orcidlink{0000-0003-1332-0006}$^{1}$ and Santabrata Das\,\orcidlink{0000-0003-4399-5047}$^{1}$}

\affil{$^1$Department of Physics, Indian Institute of Technology Guwahati, Assam-781039, India}


\begin{abstract}
The interplay between supermassive black holes (SMBHs) and their surrounding environment is fundamental to understanding galactic evolution. This work investigates the influence of a cold dark matter (DM) halo on the dynamics of relativistic, low angular momentum, inviscid, and advective hot accretion flow onto a galactic SMBH. Modeling the spacetime geometry as a black hole embedded within various DM distributions, including those with a central density spike, we demonstrate that the presence of a DM halo, particularly one that is massive and compact, enhances the luminosity of the accretion disk. The dominant contribution to this luminosity originates from the inner regions of the flow, suggesting that luminosity measurements could serve as a valuable observational probe for the dense DM environments expected near galactic centers.
\end{abstract}

\section{Introduction}
Supermassive black holes (SMBHs), with mass exceeding $10^5 M_{\odot}$, are now understood to be ubiquitous at the centers of most galaxies. Observational evidences, from the velocity dispersion of S-stars orbiting Sagittarius A*~\cite{Genzel2010, 2008ApJ...689.1044G} to the images produced by the Event Horizon Telescope (EHT)~\cite{EventHorizonTelescope:2022wkp}, has firmly established their existence. These massive objects are not isolated; they are deeply embedded within vast halos of dark matter (DM), which constitutes over 85\% of the universe's total matter density and whose presence is strongly inferred from the flat rotation curves of galaxies.

The immense gravity of an SMBH is expected to gravitationally redistribute the surrounding cold DM, leading to the formation of a significant overdensity, or a "spike," in the DM profile near the event horizon. This modified spacetime geometry provides a unique laboratory for studying strong gravity phenomena. Since SMBHs grow primarily through the accretion of baryonic matter, the properties of this infalling material are directly influenced by the underlying gravitational potential. We explore this connection by examining how a DM halo influences an advection-dominated accretion flow (ADAF), a model that accurately describes the hot, radiatively inefficient flows observed in low-luminosity active galactic nuclei (AGNs). By analyzing these effects, we can identify potential observational signatures of the DM spike~\cite{Chowdhury:2025tpt}.

\section{Formalism and Methodology}

\subsection{Spacetime Geometry with Dark Matter}
We model the spacetime geometry of a BH by the line element,
\begin{equation}\label{eq:metric}
			ds^2 =  -f(r) dt^2 + \frac{dr^2}{1-\frac{2m(r)}{r}}+r^2d\Omega^2~,
\end{equation}
where $d\Omega^2$ represents the metric on a unit two-sphere. 

Following~\cite{Cardoso:2021wlq, Figueiredo:2023gas, Speeney:2024mas, Chakraborty:2024gcr, Pezzella:2024tkf}, we employ a generalized Einstein cluster formalism~\cite{Einstein:1939ms, Geralico:2012jt}, that considers particles in all possible circular geodesics and provides an average stress tensor that represents an anisotropic fluid with only tangential pressure $P_t$. Thus, we assume the metric in \cref{eq:metric} to be a solution of the Einstein's equations,
\begin{equation}\label{eq:EFE}
	G_{\mu\nu}=8\pi T^\text {env}_{\mu\nu},
\quad \mbox{where,}\quad
\left({T}^\mu_{\nu}\right)^{\rm env}= {\rm diag}(-\rho_{{\rm DM}} (r),0,P_t(r), P_t(r))~,
\end{equation}
is the anisotropic stress-energy tensor encoding the properties of the environment in terms of the density $\rho_{{\rm DM}}(r)$ and the tangential pressure $P_t(r)$ of the matter distribution. The anisotropy of the DM distribution is due to the negligible accretion rate of the DM particles onto the central BH. For a given density profile, the continuity equation determines the mass profile as 
\begin{equation}\label{eq:mprime}
  m'(r)=4 \pi r^2 \rho_{\rm DM}(r)~,  
\end{equation}
whereas the metric function $f(r) $ and the tangential pressure $P_t (r)$ are determined by the $rr$ component of the field equations  (\cref{eq:EFE}) and the Bianchi identities, respectively,
\begin{align}   
    \frac{f'(r)}{f(r)}= \frac{2 m(r)/r}{r-2 m(r)}~,\quad
    P_t(r)=\frac{m(r)/2}{r-2 m(r)}\rho_{\rm DM}(r)~.
\end{align}
The metric functions $f(r)$ and $m(r)$ completely specify the geodesic structure of the spacetime.
In this work, we consider several DM profiles: Hernquist (HQ), Navarro-Frenk-White (NFW), and Einasto models, along with a relativistic HQ-type spike model.

Since there are multiple length scales involved in the problem, we maintain a strict separation of the length scales, $M_{\rm BH}\leq M_{{\rm halo}}\leq a_0$ and $r_{\rm edge}\leq a_0$, where $M_{\rm BH}$ is the BH mass, $M_{{\rm halo}}$ is the total mass of the DM contained in the halo, $a_0$ is the scale radius associated with the DM distribution and $r_{\rm edge}$ is the outer edge of the accretion disk. Note that in all the cases, we treat $M_{\rm halo}$ and $a_0$ as free model parameters for a qualitative analysis.
\begin{figure}[h!]
\centering
\includegraphics[width=0.6\textwidth]{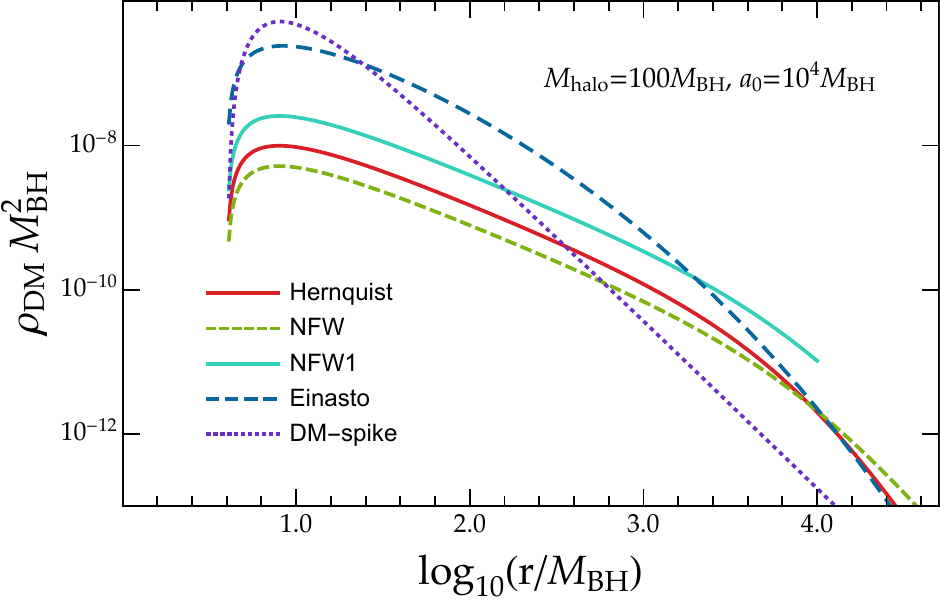} 
\caption{A plot showing the different DM density profiles (Hernquist, NFW, Einasto, Spike) as a function of radial distance. The spike model shows a significant density enhancement close to the black hole. See~\cite{Chowdhury:2025tpt} for details.}
\label{fig:dm_profiles}
\end{figure}
\subsection{Hydrodynamics of the Accretion Flow}

The accretion flow is treated as an ideal relativistic fluid, whose dynamics are governed by the conservation of the mass-flux and the fluid stress-energy tensor written as,
\begin{equation}
\nabla_{\mu}(\rho u^{\mu}) = 0; \quad \nabla_{\mu}T^{\mu\nu} = 0
\label{eq:conservation_laws}
\end{equation}
where $T^{\mu\nu} = (e+p)u^{\mu}u^{\nu} + pg^{\mu\nu}$ is the stress-energy tensor, with $e$ and $p$ being the local energy density and pressure of the fluid of density $\rho$ moving with a four velocity $u^\mu$. Since the accretion flow
is thermally relativistic, we use the relativistic equation of state (REoS) ~\cite{Chattopadhyay:2008xd,Chowdhury:2025tpt} with variable adiabatic index ($\Gamma$) to relate the density and pressure of the fluid. 

For a steady, axisymmetric flow in the equatorial plane, we derive the governing equations for the radial velocity and temperature in the corotating fluid frame. In transonic accretion, the flow starts subsonically from the outer edge of the accretion disk and reaches the black hole horizon with speed approaching that of light. A smooth accretion flow must pass through a critical point $\left(r_c\right)$, defined as the location where the numerator and denominator of the radial velocity gradient vanish simultaneously. The number and location of the critical point play a crucial role in classifying the flow topology.

\section{Results and Discussion}

The presence of a DM halo modifies the effective potential governing the accretion flow, leading to distinct observational signatures. It shifts the critical point of the accretion flow closer to the black hole. The parameter space of specific energy and specific angular momentum ($\lambda$) that permits physically valid transonic solutions is also altered. 
We observe that the gravitational effect of the DM halo reduces the fluid's inflow velocity while increasing its local temperature, resulting in a thicker, quasi-spherical disk. The most direct observational consequence is the enhancement of the disk's luminosity. We calculate the spectral energy distribution and bolometric luminosity from thermal bremsstrahlung emission, incorporating relativistic effects and electron-electron emission in addition to the standard electron-ion emission channel. Our results show that even for a give DM halo, both the mass ($M_{\rm halo}$) and the compactness of the DM halo significantly boost the luminosity

\begin{figure}[h!]
\centering
\includegraphics[width=0.75\textwidth]{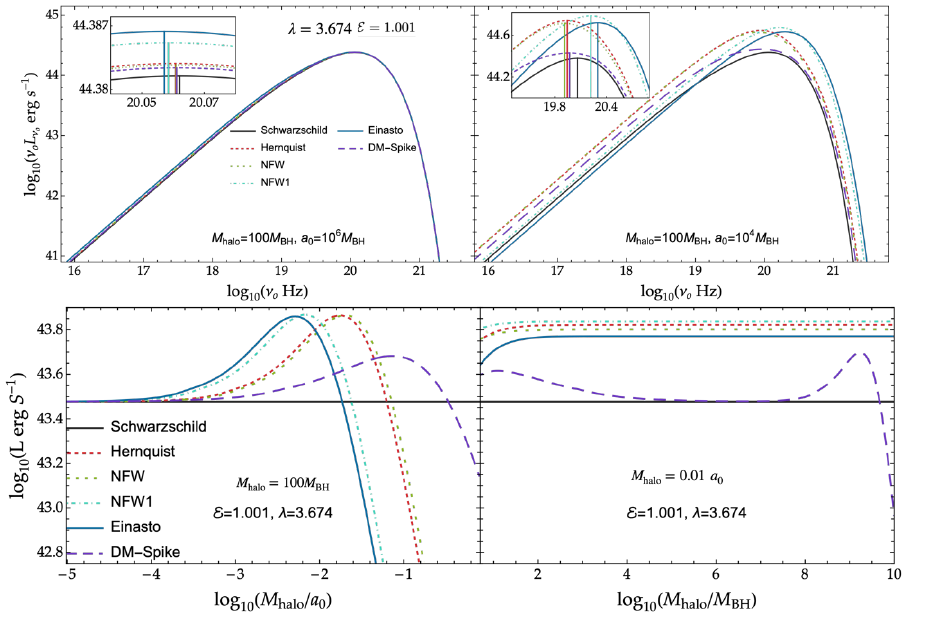} 
\caption{Top: plot of the spectral energy distribution for different DM distributions for different values scale radius: $ M_{\rm halo}=100 M_{\rm BH}, a_0=10^6 M_{\rm BH}$ (left), $ M_{\rm halo}=100 M_{\rm BH} \mbox{\ and \ } a_0=10^4 M_{\rm BH}$ (right).  Halo compactness $(M_{\rm halo}/a_0)$ changes from $10^{-4}$ (left) to $10^{-2}$ (right) for a fixed halo mass. The solid black curve in each plot shows the SED for a Schwarzschild BH of the same mass. The insets show the location of the peak of the SEDs in each case. The flow parameters are $\lambda=3.674$ and $\mathcal{E}=1.001$. Bottom: plot of bolometric luminosity for different dark matter profiles with halo compactness, $M_{\rm halo}/a_0$ for $M_{\rm halo}=100 M_{\rm BH}$ for $\mathcal{E}=1.001$ and $\lambda=3.674$ (left). Plot of the luminosity for different dark matter profiles with $M_{\rm halo}$ for $M_{\rm halo}/ a_0=0.01$ (right) for the same values of energy and specific angular momentum of the flow.}
\label{fig:luminosity}
\end{figure}
Crucially, the dominant contribution to this enhanced emission comes from the inner regions of the disk, where the DM spike's gravitational influence is strongest. 

The connection between the disk luminosity and the DM halo properties establishes a novel avenue for probing the nature of DM in strong-gravity regimes. Future high-resolution observations could test these theoretical predictions and potentially uncover the signature of a dark matter spike at the heart of our galaxy and others. Our results suggest a potential modification to our understanding of the coevolution of SMBHs and host galaxies~\cite{Kormendy:2013dxa}, though we acknowledge that degeneracies with other astrophysical factors like black hole spin and magnetic fields exist~\cite{Shapiro1974, McKinney-Gammie2004, Li-etal2004, Lynden-Bell1969, Shapiro1973, Chakrabarti-Mandal2006, Sarkar-Das2018, Shakura-Sunyaev1973, 1980A&A....88...23P, Chakrabarti-Titarchuk1995} and that a more realistic accretion model is required for future investigations.

\section*{Acknowledgments}
AC acknowledges travel support from Anusandhan National Research Foundation (ANRF), Govt. of India via the International Travel Support  (ITS) scheme to attend the 24th International Conference on General Relativity and Gravitation (GR24) and the 16th Edoardo Amaldi Conference on Gravitational Waves (Amaldi16) in Glasgow, UK. The work of AC is supported by the National Postdoctoral Fellowship of ANRF, Govt. of India (File No.: PDF/2023/000550). SC acknowledges support of the MATRICS research grant awarded by ANRF, Govt. of India, through grant no. MTR/2022/000318.

\bibliographystyle{iopart-num}
\bibliography{DM_Accretion/ref}
\end{document}